\documentstyle[preprint,aps]{revtex}

\begin{document}
\draft
\author{W. Zuo$^{1,2,3}$, Z.H. Li$^{1,2}$, A. Li$^{3}$, G.C. Lu$^{1,2}$}
\address{$^1$Institute of Modern Physics, Chinese Academy of Sciences,
  Lanzhou 730000, P.R. China\\
 $^2$Graduate School of Chinese Academy of Sciences, Beijing 100039, P.~R.~China\\
 $^3$School of Physics and Technology, Lanzhou University, Lanzhou 730000, P.~R.~China}
\title{Hot Nuclear Matter Equation of State with a Three-body Force}
\maketitle

\begin{abstract}
The finite temperature Brueckner-Hartree-Fock approach is extended
by introducing a microscopic three-body force. In the framework of
the extended model, the equation of state of hot asymmetric
nuclear matter and its isospin dependence have been investigated.
The critical temperature of liquid-gas phase transition for
symmetric nuclear matter has been calculated and compared with
other predictions. It turns out that the three-body force gives a
repulsive contribution to the equation of state which is stronger
at higher density and as a consequence reduces the critical
temperature of liquid-gas phase transition. The calculated energy
per nucleon of hot asymmetric nuclear matter is shown to satisfy a
simple quadratic dependence on asymmetric parameter $\beta$ as in
the zero-temperature case. The symmetry energy and its density
dependence have been obtained and discussed. Our results show that
the three-body force affects strongly the high-density behavior of
the symmetry energy and makes the symmetry energy more sensitive
to the variation of temperature. The temperature dependence and
the isospin dependence of other physical quantities, such as the
proton and neutron single particle potentials and effective masses
are also studied. Due to the additional repulsion produced by the
three-body force contribution, the proton and neutron single
particle potentials are correspondingly enhanced as similar to the
zero-temperature case.

\end{abstract}

\pacs{PACS numbers: 21.65.+f, 13.75.Cs, 24.10.Cn, 05.70.Ce}

\newpage

\section{Introduction}

One of the most important motivations of heavy ion physics is to
investigate the properties of nuclear matter under extreme
conditions of temperature, density and isospin. The equation of
state (EOS) of asymmetric nuclear matter at finite temperature
plays an important role in understanding the dynamics of heavy ion
collisions at intermediate and high energies. Theoretical
simulations\cite{BAO1} indicate that the main reaction dynamics
and observables including collective flow, balance energy, isospin
equilibrium, pre-equilibrium nucleon emission, isotopic scaling
and isospin diffusion are quite sensitive to the isospin dependent
part of the nuclear EOS, especially to the density dependence of
the symmetry energy. The properties of hot and dense asymmetric
nuclear matter are also of great interest in astrophysics
especially in connection with the dynamics of supernova explosions
and the thermal evolution of neutron stars\cite{PRAK}. The
temperature and isospin dependence of the EOS of asymmetric
nuclear matter is crucial for a reliable description of the
thermodynamical evolution and the structure of the newborn neutron
star ( protoneutron star ) formed in the latest stage of a type-II
supernova collapse\cite{PRAK}. Motivated by the general
significance in heavy ion physics and astrophysics, the
thermodynamical properties of nuclear matter has been investigated
extensively within various theoretical models such as the
non-relativistic Hartree-Fock approach\cite{LATT1} and the
relativistic mean field theory (RMT)\cite{WALE,MULL}.

Due to the Van der Waals nature of the nucleon-nucleon (NN)
interaction, it is generally expected that nuclear matter is
likely to exhibit a liquid-gas phase transition\cite{BERT}.
Theoretically, considerable effort has been devoted to
establishing the EOS of hot symmetric nuclear matter and
discussing the critical
phenomena\cite{KUPE,SATP,JAQA1,BALD1,JAQA2,HAAR,HUB}. Although
almost all the predicted EOSs of infinite symmetric nuclear matter
display a typical Van der Waals behavior, the obtained values of
the critical temperature $T_C$ for the phase transition are
distributed in a wide range from 8MeV to 20MeV. Within the
non-relativistic framework, the calculated value of $T_C$ for
infinite symmetric nuclear matter is $15-20$MeV\cite{JAQA1,BALD1}
depending on the choice of the theoretical models and the NN
interactions. The RMT theory gives a critical temperature of about
$14$MeV\cite{MULL}. The value of $T_C$ predicted from the
Dirac-Brueckner ( DB ) approach is as low as $\sim8$MeV in
Ref.\cite{HAAR} and about 10MeV in Ref.\cite{HUB}.

On a microscopic basis, the properties of hot asymmetric nuclear
matter have been studied in Ref.\cite{BOMB1} in the framework of
the finite temperature Brueckner-Hartree-Fock ( FTBHF ) approach
using the separable version of the Paris two-body force. As well
known, three-body forces are necessary for reproducing the
empirical saturation properties of cold symmetric nuclear matter
in a non-relativistic microscopic
approach~\cite{GRAN,LEJ1,ZUO1,BALD2,MAC}. Two kinds of three-body
force ( TBF ) have been adopted in
 the Brueckner-Hartree-Fock ( BHF )
 formalism\cite{GRAN,LEJ1,ZUO1,BALD2}. One is the
semi-phenomenological TBF\cite{PUDL} which has two or few
adjustable parameters determined by fitting the empirical
saturation density and energy of the cold symmetric nuclear matter
in the BHF calculations\cite{BALD2}. The other is the microscopic
TBF based on meson exchange coupled to the intermediate virtual
excitations of nucleon-antinucleon pairs and nucleon
resonances\cite{GRAN,LEJ1,ZUO1}.

In the present paper, our aim is to extend the FTBHF approach by
introducing the microscopic TBF and to investigate the properties
of hot asymmetric nuclear matter.
 Special attention has been payed on the effects of the TBF
 contribution and the temperature dependence. The two-body
 realistic NN interaction adopted in the present calculations is
 the Argonne $V_{18}$ ( $AV_{18}$ ) potential\cite{WIRI}.
 Our calculations indicate that the TBF affects not
only the EOS of hot asymmetric nuclear matter and the critical
temperature of the liquid-gas phase transition, but also the
single-particle ( s.p. )
 properties such as the proton and neutron s.p. potentials and
effective masses. The present paper is organized as follows. We
will describe briefly the model in Section II. The numerical
results are presented and discussed in Section III. In Section IV
a summary of the present work is given.

\section{Theoretical Model}

In general, three parameters are required to specify a given
thermodynamical state of hot asymmetric nuclear matter, i.e., the
total nucleon number density $\rho$, the isospin asymmetry
parameter $\beta =\frac{\rho_n-\rho_p}{\rho}$ and the temperature
$T$. At zero-temperature, the proton and the neutron Fermi momenta
are given by $k_F^\tau =[3\pi ^2\rho _\tau ]^{1/3}$ with $\tau =p
$ or $\tau =n$, where $\rho_n=\frac 12$ $(1+\beta )\rho $ and
$\rho_p=\frac{1}{2}(1-\beta )\rho $ are the proton and neutron
number densities, respectively. The formalism of the
Brueckner-Bethe-Goldstone (BBG) theory for cold asymmetric nuclear
matter can be found in Ref.\cite{BOMB2}. The extension to finite
temperature is given in Ref.\cite{BOMB1}. In the following, we
give a brief review for completeness.
 The starting point of the BBG
approach is the Brueckner reaction matrix $G$ which satisfies the
following generalized Bethe-Goldstone equation,
\begin{equation}\label{e:ftbg}
G_{\tau ,\tau ^{\prime }}(\rho ,\beta, T, \omega )=v+v\sum\limits_{k_1k_2}%
\frac{\mid k_1k_2\rangle Q_{\tau ,\tau ^{\prime }}\langle k_1k_2\mid }{%
\omega -e_{\tau}(k_1)-e_{\tau'}(k_2)}G_{\tau ,\tau ^{\prime
}}(\rho, \beta, T,\omega )
\end{equation}
where $v=v_2+V_3^{\rm eff}$ is the nucleon-nucleon (NN)
interaction and $\omega $ is the starting energy. In the present
calculations the Argonne $V_{18}$ ($AV_{18}$) potential\cite{WIRI}
is adopted as the bare two-body force $v_2$ and $V_3^{\rm eff}$ is
the TBF contribution given by Eq.(\ref{e:tbf})(see below). The
finite temperature Pauli operator $Q_{\tau,\tau ^{\prime }}$ is
simply an extension of the zero-temperature one, i.e,
\begin{equation}\label{e:fq}
Q_{\tau ,\tau ^{\prime }}=Q_{\tau ,\tau ^{\prime }}(k_1, k_2,\rho,
\beta,T)=[1-f_\tau (k_1,\rho ,\beta,T )][1-f_{\tau ^{\prime
}}(k_2,\rho ,\beta, T)]
\end{equation}
At finite temperature the Fermi distribution is expressed as,
\begin{equation}
f_\tau (k,\rho,\beta,T)=\left[1+\exp \left(\frac{e_\tau (k)-\mu
_\tau }T\right)\right]^{-1}
\end{equation}
where $\mu _\tau =\mu _\tau (k,\rho,T)$ refers to the chemical
potential and $e_\tau (k)$ is the s.p. energy. For any given
density and temperature, we can calculate the chemical potential
$\mu _\tau $ from the following implicit equation
self-consistently by iteration,
\begin{equation}\label{e:den}
\rho _\tau =\frac 1V\sum_kf_\tau (k,\rho,\beta,T)=\frac 1V\sum_k
\left[1+\exp \left(%
\frac{e_\tau (k)-\mu _\tau }T\right)\right]^{-1}
\end{equation}
The s.p. energy is given by,
\begin{equation}\label{e:spe}
e_\tau (k)\equiv e_\tau (k,\rho ,\beta,T)=\frac{\hbar ^2k^2}{2m}+
U_\tau (k, \rho, \beta, T)
\end{equation}
where $U_\tau (k,\rho ,\beta, T)$ is the proton or neutron s.p.
potential. Within the FTBHF framework the s.p. potential is
calculated from the on-shell anti-symmetrized $G$ matrix,
\begin{eqnarray}\label{e:spu}
U_\tau (k,\rho, \beta, T) = \frac{1}{2} \sum_{\tau^{\prime}}
\sum_{\vec k^{\prime}} f_{\tau^{\prime}}(k^{\prime},\rho,\beta, T)
\langle kk^{\prime} \mid G_{\tau,\tau^{\prime }}(\rho, \beta, T,
e_{\tau}(k)+e_{\tau^{\prime}}(k^{\prime}))\mid
kk^{\prime}\rangle_A
\end{eqnarray}
In the present calculations, we adopt the continuous choice
\cite{JEUK} for the s.p. potential. On the one hand, the
continuous choice has been shown to provide a much faster
convergence of the hole-line expansion for the energy per nucleon
in nuclear matter at $T=0$ than the gap choice~\cite{SONG}. On the
other hand, in the continuous choice, the s.p. potential describes
physically the nuclear mean field felt by a nucleon in nuclear
medium.


The microscopic TBF adopted in the present calculations is
constructed from the meson-exchange current approach~\cite{GRAN}.
Its components are depicted diagrammatically in Fig.~1,  taken
from Ref.\cite{GRAN}. Four important mesons $\pi $, $\rho $, $
\sigma $ and $\omega $ are considered\cite{MAC}. The TBF contains
the contribution of the two-meson exchange part of the NN
interaction medium-modified by the intermediate virtual excitation
of nucleon resonances, the term associated to the non-linear
meson-nucleon coupling required by the chiral symmetry, the
simplest contribution rising from meson-meson interaction and
finally, the two-meson exchange diagram with the virtual
excitations of nucleon-antinucleon pairs. The meson masses in the
TBF have been fixed at their physical values except for the
virtual $\sigma$-meson mass which has been fixed at $540$MeV
according to Ref.~\cite{GRAN}. This value has been checked to
satisfactorily reproduce the $AV_{18}$ interaction from the
one-boson-exchange potential (OBEP) model\cite{ZUO1}. The other
parameters of the TBF, i.e., the coupling constants and the
regular masses (form factors), have been determined from the OBEP
model to meet the self-consistent requirement with the adopted
$AV_{18}$ two-body force. For a more detailed description of the
model and of the approximations we refer to
Refs.~\cite{GRAN,LEJ1,ZUO1}.

The TBF effect is included in the self-consistent Brueckner
procedure along a similar way to the zero temperature
case\cite{GRAN,LEJ1}, where an effective two-body interaction is
constructed to avoid the full three-body problem. A detailed
description and justification of the procedure can be found in
Ref.\cite{GRAN}. Being Extended to finite temperature, the
effective two-body interaction $V_3^{\rm eff}(T)$ in $r$-space is
given by,
\begin{eqnarray}\label{e:tbf}
 \langle \vec r_1^{\ \prime} \vec r_2^{\ \prime}| V_3^{\rm eff} (T)|
\vec r_1 \vec r_2 \rangle &=& \displaystyle
 \frac{1}{4} {\rm Tr} \sum_{k_n} f(k_n,\rho,\beta,T) \int {\rm d}
{\vec r_3} {\rm d} {\vec r_3^{\ \prime}}\phi^*_n(\vec r_3^{\
\prime}) (1-\eta(r_{13}', T )) \displaystyle (1-\eta(r_{23}', T))\nonumber \\[2mm]
&\times&
 W_3(\vec r_1^{\ \prime}\vec r_2^{\ \prime} \vec r_3^{\ \prime}|\vec
r_1 \vec r_2 \vec r_3) \phi_n(r_3) (1-\eta(r_{13}, T))
 (1-\eta(r_{23}, T))
\end{eqnarray}
where the trace is taken with respect to the spin and isospin of
the third nucleon. The function $\eta(r,T)$ is the correlation
function ( or called defect function )\cite{LEJ2}. The defect
function is directly related to the solution of the
finite-temperature Bethe-Goldstone equation, i.e.
eq.(\ref{e:ftbg})\cite{JEUK,LEJ2} and depends on temperature. The
transformation of the TBF to the above effective interaction
entails a self-consistent coupling between the TBF and the
Brueckner procedure of solving the Bethe-Goldstone equation. One
first calculates the correlation function with only the two-body
force and then builds up the effective interaction $V_3^{\rm
eff}(T)$ which in turn is added to the bare two-body force, and
again evaluates the correlation function and so on up to
convergence is reached.

It is worth to noticing that the TBF itself is the same as that
adopted in the zero-temperature case\cite{ZUO1} and independent of
temperature. As compared to the zero-temperature case, the
effective two-body force $V_3^{\rm eff}(T)$ constructed from the
TBF depends on temperature due to the medium effects and
consequently its effects are expected to be more pronounced at
finite temperature. It is obviously from Eq.(\ref{e:tbf}) that the
temperature dependence of $V_3^{\rm eff}(T)$ comes from the Fermi
distribution $f(k,\rho,\beta,T)$ and the defect function
$\eta(r,T)$. This temperature dependence is somewhat similar to
that of the Skyrme force where a three-body part is equivalent to
a two-body density-dependent effective interaction which depends
implicitly on temperature via density\cite{JAQA2}.

We solve Eqs.(\ref{e:ftbg}), (\ref{e:den}), (\ref{e:spe}),
(\ref{e:spu}), (\ref{e:tbf}) self-consistently to get the
$G$-matrix for any given density, temperature and isospin
asymmetry. In general, five iterations is needed to reach a
satisfactory convergence. The total energy $E(\rho,\beta,T
)=E_{kin}(\rho ,\beta,T)+E_{pot}(\rho,\beta,T)$ can be obtained
from the $G$-matrix. The total kinetic energy is
\begin{equation}
E_{kin}(\rho,\beta,T)=\sum_\tau \sum_kf_\tau
(k,\rho,\beta,T)\frac{\hbar ^2k^2}{2m}
\end{equation}
and the total potential energy
\begin{eqnarray}
E_{pot}(\rho,\beta,T) =\frac 12\sum_{\tau,\tau ^{\prime
}}\sum_{k,k^{\prime}}f_\tau (k,\rho,\beta,T)f_{\tau ^{\prime
}}(k^{\prime},\rho,\beta,T) \langle kk^{\prime}\mid G_{\tau,\tau
^{\prime }}(\rho,\beta,T,e(k)+e(k^{\prime}))\mid kk^{\prime
}\rangle_A
\end{eqnarray}
The total entropy $S$ can be evaluated in the approximation of a
non-interacting Fermi gas of quasi-particle in the mean field
$U_\tau (k,\rho,\beta,T)$\cite{BOMB1}. The free energy and the
pressure can be obtained according to the standard thermodynamic
relation, i.e,
\begin{equation}
F=E-TS
\end{equation}
\begin{equation}
P=\rho ^2\left(\frac{\partial F}{\partial \rho }\right)_{T,\beta}
\end{equation}
Based on the above extended FTBHF approach including the TBF, we
can investigate the thermodynamic properties of hot asymmetric
nuclear matter at various values of density, temperature and
asymmetry.

\section{Numerical Results and Discussions}

 The EOS of symmetric
nuclear matter at finite temperature is reported in Fig.2, where
the isotherms of pressure versus density is given for six values
of temperature $T=0,8,10,12,14,16$MeV from the bottom to the top.
In the figure the solid and dashed curves indicate the results
obtained by using the $AV_{18}$ two-body force plus the TBF and
the pure $AV_{18}$ two-body force, respectively. As expected, the
TBF gives a repulsive contribution to the EOS of nuclear matter.
This contribution becomes stronger as increasing density and makes
the EOS at high density much stiffer. At zero temperature, the
additional repulsion from the TBF leads to a decisive improvement
of the calculated saturation density of symmetric nuclear matter
towards the empirical value ( comparing the lowest solid and
dashed curves ). It is clearly seen from the figure that in both
cases with and without the TBF, the predicted EOS exhibits a
typical Van der Waals structure in agreement with the results from
the Skyrme-Hartree-Fock calculations\cite{JAQA1} and from the
FTBHF approach based on two-body NN interactions\cite{BALD3},
which indicates that the infinite nuclear matter can undergo a
liquid-gas phase transition. Without the TBF, the critical
temperature $T_C$ of the liquid-gas transition is approximately
$16$MeV. Inclusion of the TBF reduces the critical temperature to
about $13$MeV which is smaller than the values $T_C=15-20$MeV from
the Skryme-Hartree-Fock calculations\cite{JAQA1}. The TBF
reduction of the critical temperature is readily understood since
the TBF effect becomes more pronounced as increasing density and
temperature. In finite nuclei, the inclusion of the finite size
effects and the Coulomb force may lead to a considerable reduction
of $T_C$ as discussed in Ref.\cite{JAQA2}. The present value
$T_C\simeq13$MeV is comparable with the value $T_C\simeq14$MeV
predicted by the relativistic mean field theory\cite{MULL} but it
is larger than the ones obtained from the Dirac-Brueckner ( DB )
 approach, $T_C\simeq8\sim9$MeV in Ref.\cite{HAAR} and
$T_C\simeq10$MeV in Ref.\cite{HUB}. One possible reason for this
discrepancy is the difference between the relativistic effect in
the DB approach\cite{BROW} and the effect of the present
microscopic TBF as discussed in Ref.\cite{ZUO1}.


In Fig.3 is reported the proton or neutron s.p. potential in
symmetric nuclear matter at a fixed density $\rho =0.16$fm$^{-3}$
for three different values of temperature $T=0, 10$ and $20$MeV.
As increasing temperature, the s.p. potential becomes more
repulsive, resulting in an enhancement of the potential part of
the energy per nucleon in nuclear matter at finite temperature. It
is also seen that the curvature around the Fermi momentum becomes
more smooth as increasing temperature. This is in agreement with
the prediction obtained by using only pure two-body
forces\cite{BALD3} and is attributed to the thermal excitations
around the Fermi surface at finite temperature.

Due to the isospin effect, the proton and neutron s.p. potentials
are different in asymmetric nuclear matter. In order to discuss
the TBF effects on the isospin dependence of the proton and
neutron s.p. potentials, in Fig.4 is depicted the proton and
neutron s.p. potentials at momentum $k=0$ as a function of
asymmetry parameter for $\rho =0.16$fm$^{-3}$ and $T=10$MeV. In
the figure, the solid and dashed curves indicate the results
obtained by adopting the $AV_{18}$ plus the TBF and the pure
$AV_{18}$ two-body force, respectively. It is clear from the
figure that the inclusion of the TBF in the calculations leads to
an additional repulsion for both the proton and neutron s.p.
potentials in the whole isospin asymmetry range $0\le\beta\le1$.
As similar to the zero-temperature case, the proton s.p. potential
becomes more attractive while the neutron one becomes more
repulsive going from symmetric ($\beta=0$) to neutron ($\beta=1$)
matter. Such a different isospin dependence of the proton and
neutron s.p. potentials stems mainly from the attractive
contribution of the isospin singlet $SD$ tensor channel which
becomes stronger for protons and weaker for neutrons as increasing
neutron excess\cite{BOMB2}. As compared to the two-body force
predictions, the TBF contribution shift the proton and neutron
s.p. potentials to higher values. In both cases with and without
the TBF contribution, the dependence of the proton and neutron
s.p. potentials on the asymmetry parameter $\beta$ is almost
linear as the same as in the zero-temperature case, which supports
microscopically the Lane assumption\cite{LANE} and extends its
validity to the case of finite temperature.


In heavy-ion collisions using neutron-rich nuclei, hot nuclear
matter of high isospin asymmetry could be formed and the reaction
dynamics is expected to be very sensitive to the density and
temperature dependence of the symmetry energy. In the prompt
explosion model of a type-II supernova the $\beta$-capture process
drives the star to the last
 stage of the collapse where the proton fraction is about 0.33 and
 the temperature of a few tens of MeV may be reached. Therefore,
the understanding of the cooling mechanism of a protoneutron star
requires the information of the symmetry energy in hot nuclear
matter with high accuracy\cite{PETH,LATT}.

In Fig.~5 is depicted the shift of the energy per nucleon between
asymmetric nuclear matter and symmetric nuclear matter,
$E_A(\rho,\beta,T)-E_A(\rho,\beta=0,T)$, versus $\beta^2$ at
$T=20$MeV for two different densities $\rho=0.16$ and
0.32fm$^{-3}$. It is shown that the energy per nucleon of
asymmetric nuclear matter at finite temperature fulfills a
quadratic dependence on asymmetry parameter $\beta$ in the whole
asymmetry range ( $0\le\beta\le1$ ) as the same as in the case for
zero temperature\cite{BOMB2,HUBE,BORD}. The above $\beta^2$-law
comes originally from the empirical mass formula for small isospin
asymmetry. The present calculations show that the validity of the
empirical $\beta^2$-law can be extended to the highest isospin
asymmetry and to the case of finite temperature.

Besides the $\beta^2$-dependent term, an additional binding term
called Wigner energy has been found for the $N=Z$
nuclei\cite{ZELD}. The Wigner energy is linear in $\beta$ and
manifests itself as a spike in the experimental isobaric mass
parabola. The Wigner energy has been explored and discussed
extensively in the literatures\cite{DEAN}. It is generally
expected that the Wigner energy originates from the neutron-proton
($np$) pairing correlations and is the strongest prefigure of the
$np$ paring in the $N=Z$ nuclei\cite{SATU}. In the present
calculations, we do not consider the influence of the $np$ pairing
correlations. Since the $np$ pairing goes to vanishing very
rapidly as soon as the isospin asymmetry and the temperature
deviates from zero\cite{SEDR}, we expect that the effect of the
$np$ pairing correlation shall be appreciable only for the cold
symmetric nuclear matter, which is found to slightly lower the
energy per nucleon of the cold symmetric nuclear matter by a
self-consistent treatment of the BCS gap equation and the BHF
equations\cite{LOMB2}.

The symmetry energy is defined as
\begin{equation}
E_{\rm sym} (\rho,T) = \frac{1}{2}\left[
\frac{\partial^2E_A(\rho,\beta,T)}{\partial
\beta^2}\right]_{\beta=0}
\end{equation}
Due to the simple quadratic dependence of the energy per nucleon
on asymmetry parameter, the symmetry energy can be equivalently
calculated as the difference between the energy per nucleon of
pure neutron matter and symmetric nuclear matter, i.e.,
\begin{equation}
E_{sym}(\rho,T) \,=\, E_A(\rho,\beta=1,T) - E_A(\rho,\beta=0,T).
\end{equation}

To see the influence of temperature on the symmetry energy, in the
left panel of Fig.6 we show the density dependence of the symmetry
energy for three different temperatures $T=0, 10$ and $20$MeV. It
is seen that the symmetry energy decreases as increasing
temperature and at finite temperature it remains a monotonic
increasing function of density. As compared to the result by using
only a two-body force\cite{LOMB1}, the inclusion of the TBF
contribution enlarges the sensitivity of the symmetry energy with
respect to the variation of temperature, especially at high
densities. For a given temperature, the TBF contribution leads to
a strong enhancement of the symmetry energy at high densities as
shown in the right panel of Fig.~6 where the results for $T=20$MeV
obtained by using the $AV_{18}$ plus the TBF ( solid curve ) and
without the TBF ( dashed curve ) are plotted. It is also seen that
the symmetry energy predicted by including the TBF contribution
rises much more steeply with density than the corresponding
two-body force prediction.

Before summary, we discuss briefly the temperature dependence of
the effective mass and the TBF effect on the effective mass. The
effective mass $m_{\tau}^{*}$ for neutron ($\tau={\rm n}$) or
proton $\tau={\rm p}$ is defined as\cite{BOMB1}
\begin{equation}
\frac{m_\tau^{*}(k)}m=\frac km\left(\frac{de_\tau
(k)}{dk}\right)^{-1}
\end{equation}
where $e_\tau (k)$ is the neutron or proton s.p. energy. The
effective mass describes the nonlocal part of the nuclear mean
field which makes the local part less attractive for a nucleon
travelling with momentum $k>0$. In Fig.7 is shown the momentum
dependence of $m^{*}$ for symmetric nuclear matter at a fixed
density $\rho=0.16$fm$^{-3}$ and three different values of
temperature $T=0$,$10$ and $20$MeV. The results are comparable
with the ones reported in Refs.\cite{GRANG,HASS}. As similar to
the zero-temperature case, the momentum dependence of $m^*(k)$ is
characterized by a wide bump around the Fermi momentum due to the
high probability for particle-hole excitations nearby Fermi
surface\cite{JEUK}. However as the temperature increases, the peak
of $m^*$ becomes flatter and the peak value of $m^*$ becomes
lower. These results are in agreement with the previous
calculations by adopting only two-body
forces\cite{BOMB1,LEJ2,BALD3} and related directly with the
temperature effect on the s.p. potential around the Fermi
momentum.

To see the isospin effects, in Fig.~8 we display the neutron and
proton effective masses at their respective Fermi momenta
$k_F^{\rm n}$ and $k_F^{\rm p}$ for $T=10$MeV as a function of
asymmetry parameter $\beta$. In the figure, the solid curves are
calculated by using the $AV_{18}$ plus the TBF, while the dashed
curves by adopting only the $AV_{18}$ two-body force. In both
cases with and without including the TBF contribution, the neutron
effective mass $m^*_n$ increases and the proton one $m^*_p$
decreases as increasing $\beta$. The TBF effect is to shift
$m^*_n$ and $m^*_p$ to slightly higher values. One can also see
from the figure that the linear scissor-shaped behavior predicted
for $T=0$\cite{BOMB2} remains unchanged in the case for finite
temperature.

\section{Summary and Conclusion}

In summary, we have introduced the TBF contribution into the FTBHF
approach and investigated the EOS of asymmetric nuclear matter at
finite temperature. By using the extended model, we find that the
TBF contribution to the EOS is repulsive. The TBF repulsion
increases rapidly as increasing density and consequently the
high-density EOS including the TBF contribution becomes much
stiffer as compared to the one obtained by adopting only the
two-body force. Within the extended model, the calculated EOS of
hot symmetric nuclear matter exhibits a typical Van der Waals
structure, implying the presence of a liquid-gas phase transition.
The obtained critical temperature of the liquid-gas phase
transition is roughly $16$MaV in the case without including the
TBF contribution. The TBF leads to a $3$MeV reduction of the
critical temperature to about $13$MeV which is between
$T_C=15-20$MeV from the non-relativistic approaches and
$T_C\simeq10$Mev from the Dirac-Brueckner method.

Our calculations show that the simple $\beta^2$-law fulfilled by
the energy per nucleon of asymmetric nuclear matter at $T=0$ can
be extended to the case of finite temperature. This is of
interest, since it means that the energy per nucleon of hot
asymmetric nuclear matter can be extracted from the two limiting
cases of symmetric nuclear matter and pure neutron matter. The
symmetry energy has been calculated. For a given temperature, the
symmetry energy is a monotonic increasing function of density in
both cases with and without the TBF. At a fixed density, the
symmetry energy decreases as the temperature increases. The TBF
gives a strong enhancement of the symmetry energy at high density
and makes the symmetry energy rise much more rapidly as compared
to the results without including the TBF contribution.

The temperature dependence and the isospin dependence of the
proton and neutron s.p. potentials and effective masses have also
been calculated and discussed. As expected, the s.p. potential
becomes more repulsive and the peak of the effective mass around
the Fermi momentum becomes flatter as increasing temperature. It
is shown that the neutron effective mass increases and the proton
one decreases as increasing $\beta$. Both the TBF contribution and
the finite temperature effect do not change this isospin behavior
of the neutron and proton effective masses.

Our results show that the TBF contribution affects considerably
the stiffness of the EOS of hot nuclear matter and the symmetry
energy at high density, and as a consequence, it may have
important implication for the investigations of the dynamics
evolution of supernova explosions and the structure of
protoneutron stars.

\section{Acknowledgment}

The work is supported in part by the Knowledge Innovation Project
of the Chinese Academy of Sciences ( KJCX2-SW-N02 ), the Major
State Basic Research Development Program of
 China ( G2000077400 ), and the Major Prophase Research Project of
Fundamental Research of the Ministry of Science and Technology of
China ( 2002CCB00200 ), the National Natural Science Foundation of
China ( 10235030, 10175082 ).

\newpage

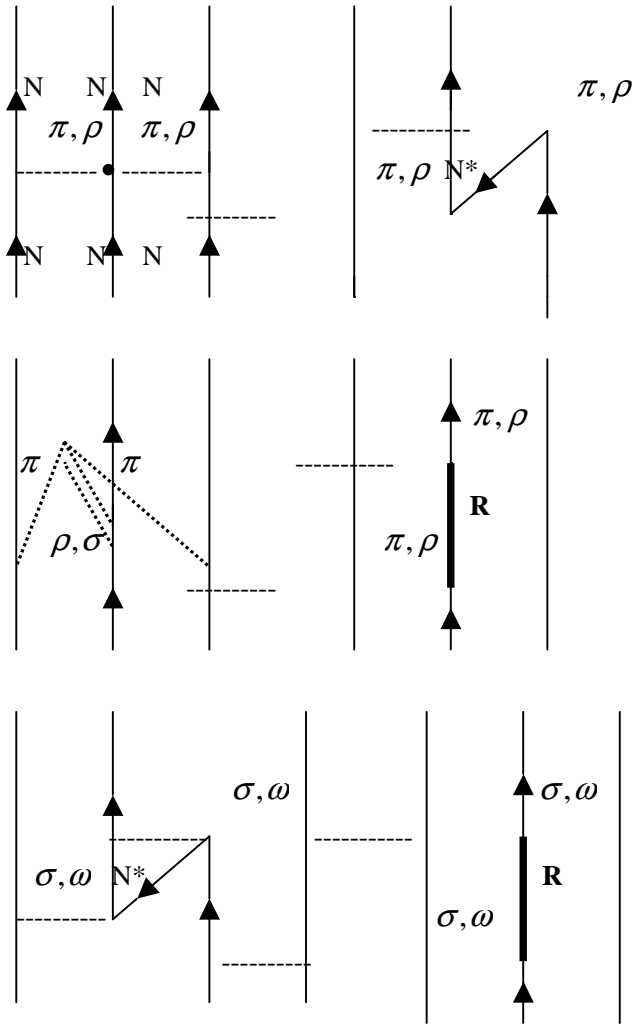
\begin{figure}[tbp]
\caption{ Diagrams of the microscopic TBF adopted for the present
calculations, taken from Ref.[15]} \label{fig1}
\end{figure}

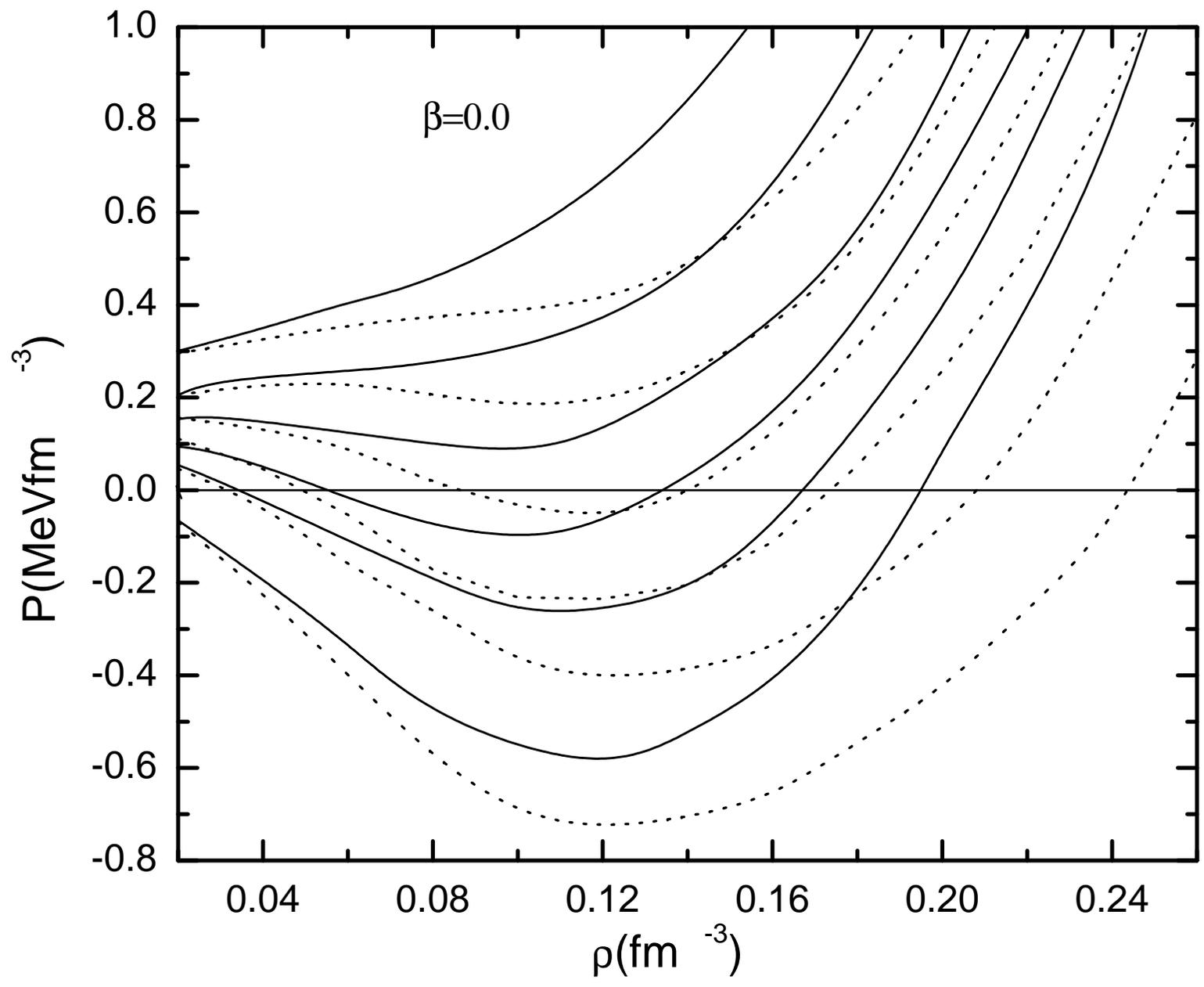
\begin{figure}[tbp]
\caption{ Pressure as function of density for symmetric matter at
six values of temperature $T=0,8,10,12,14,16$MeV from the bottom
to the top. The solid and dashed curves are the results by using
the $AV_{18}$ plus the TBF and the pure $AV_{18}$ two-body force,
respectively} \label{fig2}
\end{figure}

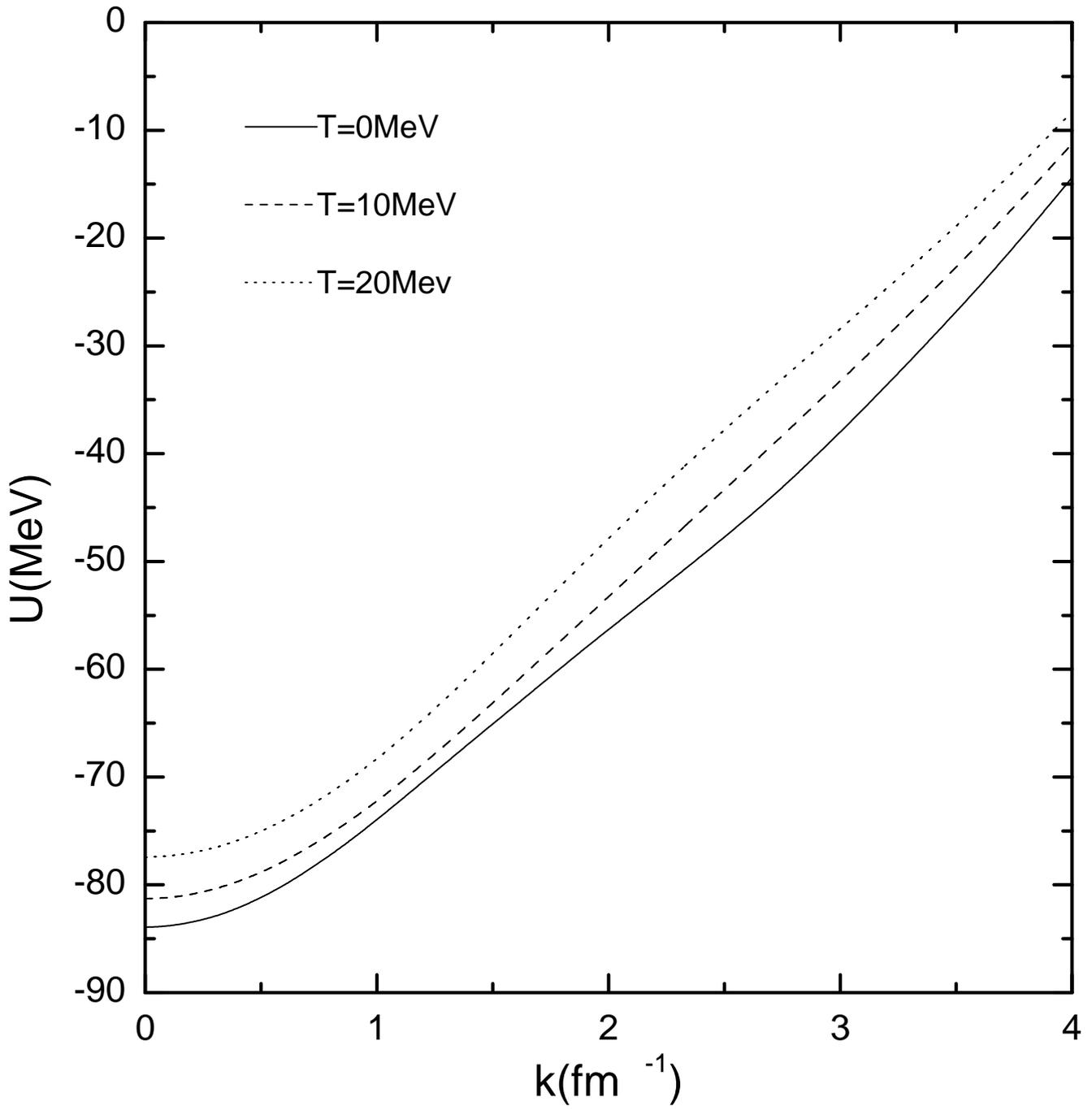
\begin{figure}[tbp]
\caption{Momentum dependence of the single-particle potential in
symmetric nuclear matter at a fixed density $\rho =0.16fm^{-3}$
and three different temperatures $T=0, 10, 20$MeV. In the
calculations, the TBF is included.} \label{fig3}
\end{figure}

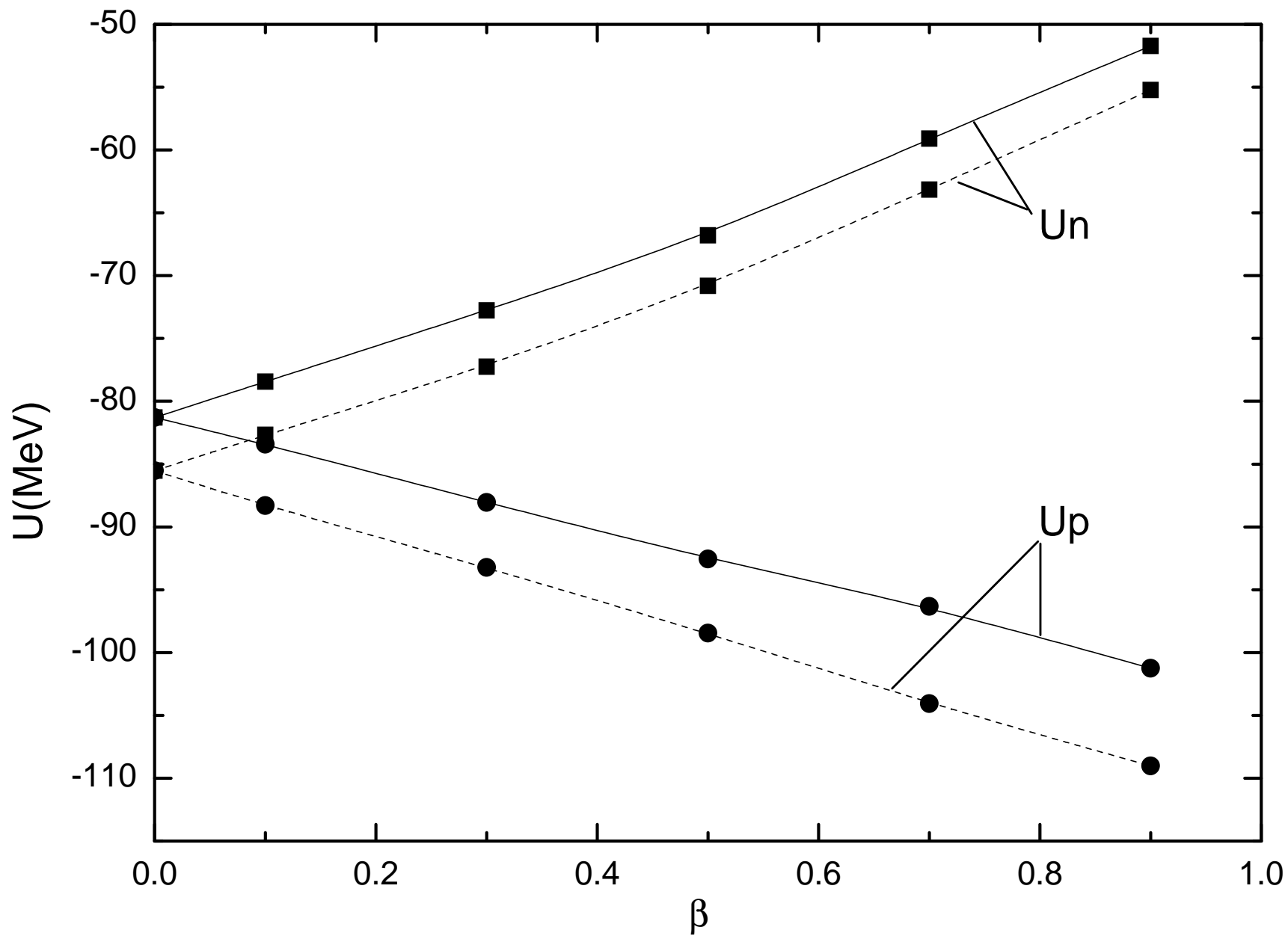
\begin{figure}[tbp]
\caption{Proton and neutron single-particle potentials at $k=0$
versus asymmetry parameter $\beta$ for $\rho=0.16$fm$^{-3}$ and
$T=10$MeV. The solid curves are the results by including the TBF
contribution and the dashed ones without the TBF contribution. }
\label{fig4}
\end{figure}

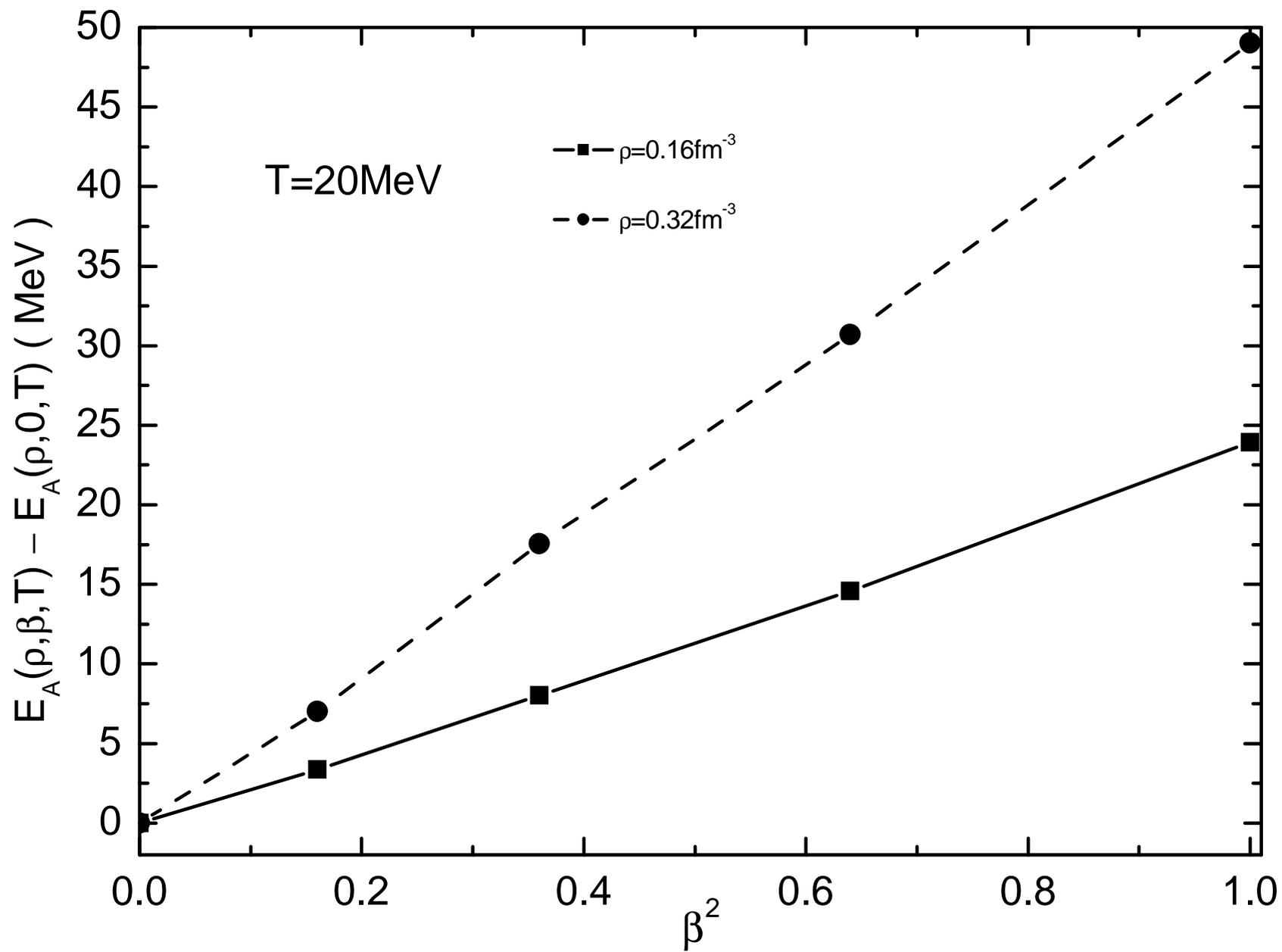
\begin{figure}[tbp]
\caption{Energy per nucleon
$E_A(\rho,\beta,T)-E_A(\rho,\beta=0,T)$ versus $\beta ^2$ in the
range $0\leq \beta \leq 1$ for two different densities. The
results are obtained by including the TBF. } \label{fig5}
\end{figure}

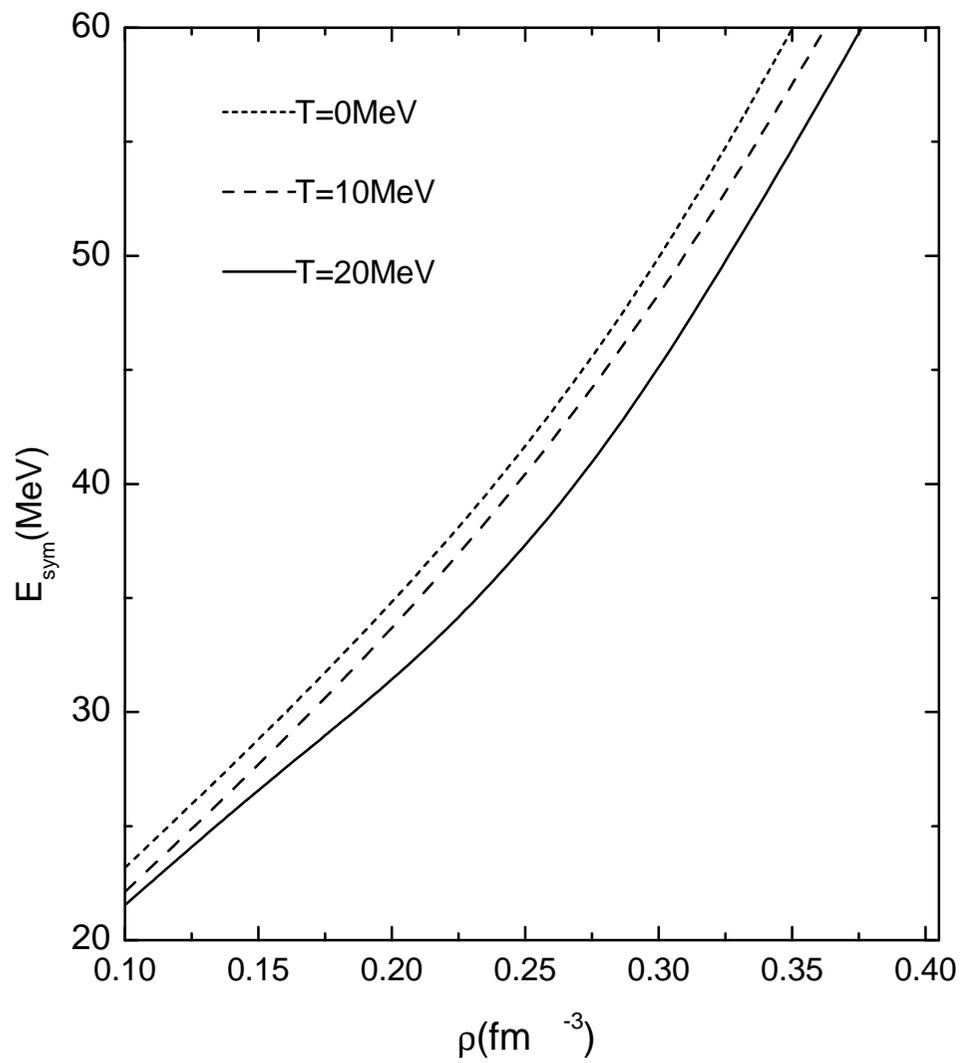
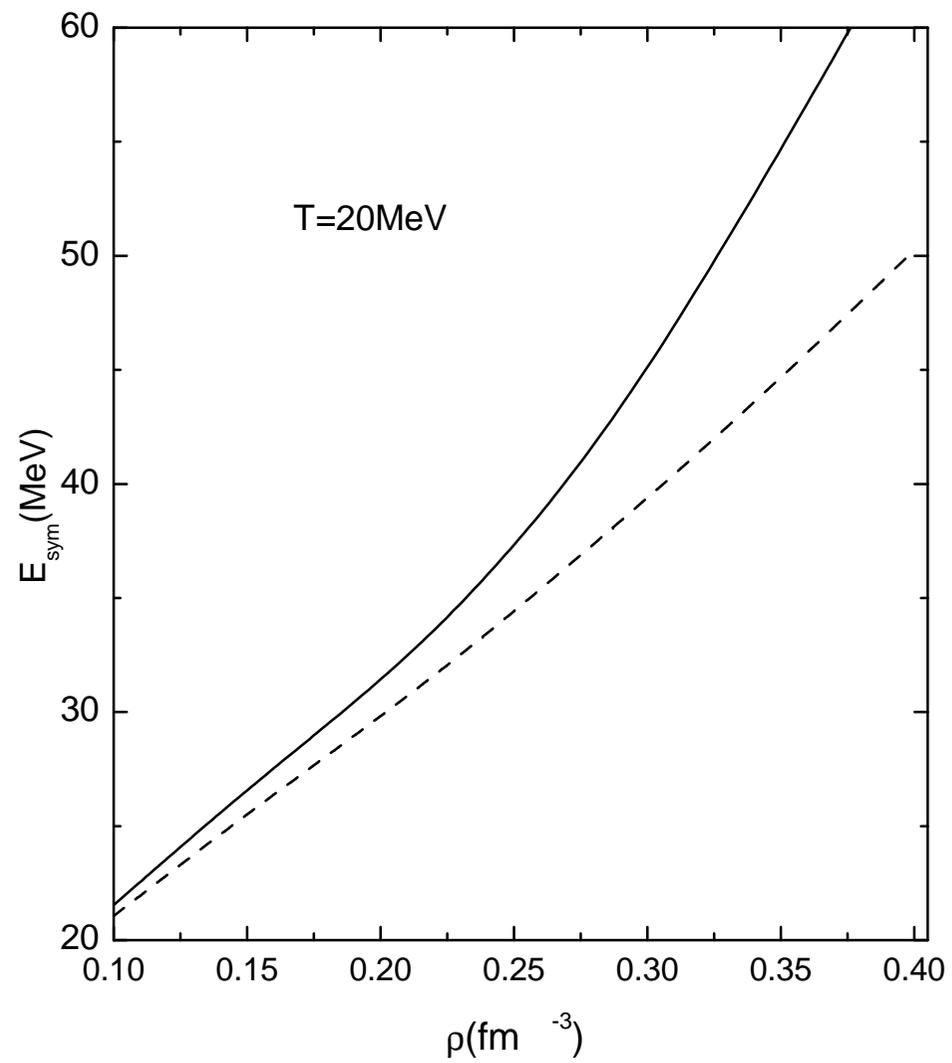
\begin{figure}[tbp]
\caption{Left panel: Symmetry energy as a function of density for
three values of temperature. The results are obtained by including
the TBF.  Right panel: Density dependence of the symmetry energy
with the TBF included (solid line) and not included (dashed line)
at a fixed temperature $T=10$MeV.} \label{fig6}
\end{figure}

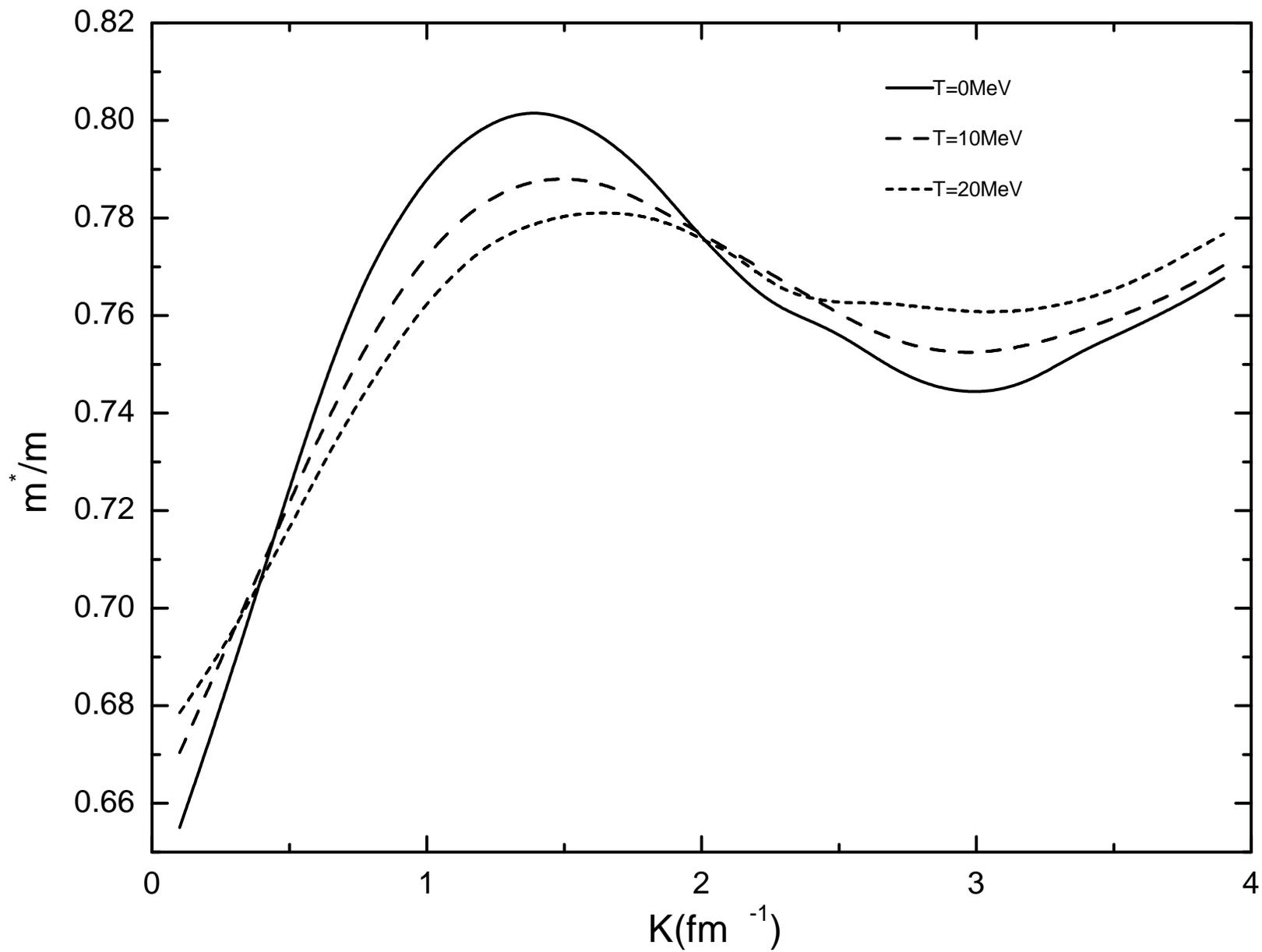
\begin{figure}[tbp]
\caption{ Momentum dependence of the nucleon effective mass in
symmetric nuclear matter at three different values of
temperature.} \label{fig7}
\end{figure}

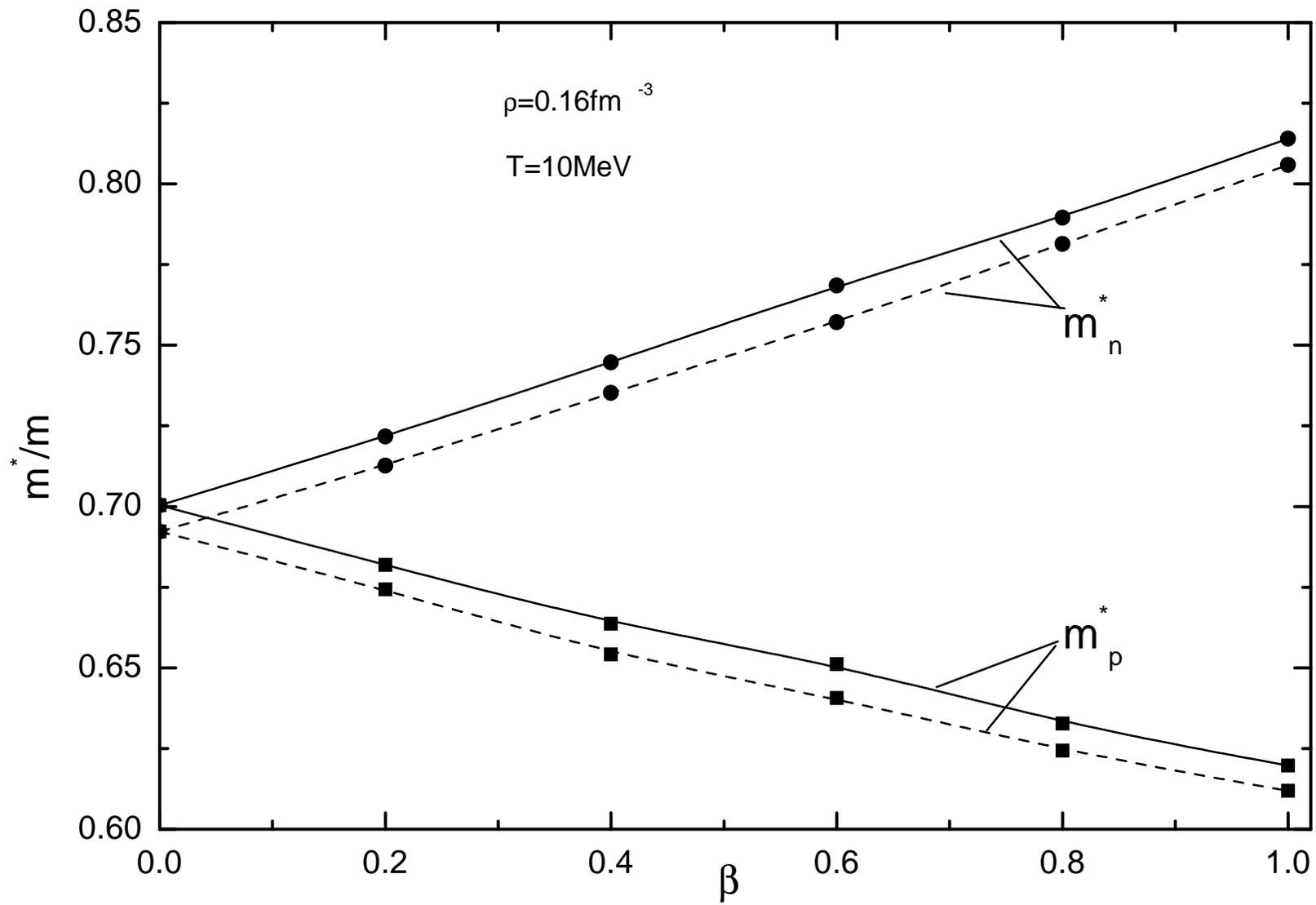
\begin{figure}[tbp]
\caption{Proton and neutron effective masses as a function of
asymmetry parameter obtained by adopting the $AV_{18}$ plus the
TBF ( solid curves ) and by using only the $AV_{18}$ two-body
force ( dashed curves ). } \label{fig8}
\end{figure}

\end{document}